\documentclass[
 reprint,
 superscriptaddress,
 amsmath,amssymb,
 aps,
 prapplied,
]{revtex4-2}

\usepackage{graphicx}
\usepackage{dcolumn}
\usepackage{bm}

\graphicspath{{./}}

\usepackage{physics}
\usepackage{siunitx}
\usepackage{acro}
\usepackage{chemformula}
\usepackage{multirow}

\newcommand{\dETwoOne}{\Delta E_{21}}
\newcommand{\omegaphoton}{\omega_{\textrm{ph}}}
\newcommand{\Ephoton}{E_{\textrm{ph}}}
\newcommand{\Vth}{V_{\mathrm{th}}}
\newcommand{\VL}{V_{\mathrm{L}}}
\newcommand{\VH}{V_{\mathrm{H}}}
\newcommand{\VGS}{V_{\mathrm{GS}}}
\newcommand{\tgate}{t_{\mathrm{gate}}}
\newcommand{\trise}{t_{\mathrm{rise}}}
\newcommand{\tfall}{t_{\mathrm{fall}}}
\newcommand{\tgatedelay}{t_{\mathrm{gate,del}}}
\newcommand{\etal}{\textit{et al.}}
\newcommand{\DOne}{\mathrm{D}_{\mathrm{1}}}
\newcommand{\ZOneTwo}{\mathrm{Z}_{\mathrm{1,2}}}
\newcommand{\EHSixSeven}{\mathrm{EH}_{\mathrm{6/7}}}
\newcommand{\EKTwo}{\mathrm{EK}_{\mathrm{2}}}
\newcommand{\FWHM}{\text{FWHM}}
\newcommand{\Egap}{E_{\textrm{gap}}}
\newcommand{\dELor}{\Delta E_{\textrm{L}}}

\DeclareAcronym{MOSFET}{short = MOSFET, long = metal-oxide-semiconductor field-effect transistor, foreign-plural={}}
\DeclareAcronym{SiC}{short = SiC, long = silicon carbide, foreign-plural={}}
\DeclareAcronym{Si}{short = \ch{Si}, long = silicon, foreign-plural={}}
\DeclareAcronym{SiO2}{short = \ch{SiO2}, long = silicon dioxide, foreign-plural={}}
\DeclareAcronym{GaN}{short = GaN, long = gallium nitride, foreign-plural={}}
\DeclareAcronym{Vth}{short = $\Vth$, long = threshold voltage, foreign-plural={}}
\DeclareAcronym{VL}{short = $\VL$, long = low level, foreign-plural={}}
\DeclareAcronym{VH}{short = $\VH$, long = high level, foreign-plural={}}
\DeclareAcronym{VGS}{short = $\VGS$, long = gate-source voltage, foreign-plural={}}
\DeclareAcronym{tgate}{short = $\tgate$, long = gate time window, foreign-plural={}}
\DeclareAcronym{tgatedelay}{short = $\tgatedelay$, long = gate delay time, foreign-plural={}}
\DeclareAcronym{trise}{short = $\trise$, long = rise time, foreign-plural={}}
\DeclareAcronym{tfall}{short = $\tfall$, long = fall time, foreign-plural={}}
\DeclareAcronym{REDR}{short = REDR, long = recombination-enhanced defect-reaction, foreign-plural={}}
\DeclareAcronym{NMP}{short = NMP, long = non-radiative multiphonon, foreign-plural={}}
\DeclareAcronym{DMOS}{short = DMOS, long = double-diffused MOS, foreign-plural={}}
\DeclareAcronym{TO}{short = TO, long = transistor outline, foreign-plural={}}
\DeclareAcronym{CCD}{short = CCD, long = charge-coupled device, foreign-plural={}}
\DeclareAcronym{RMS}{short = RMS, long = root-mean-square, foreign-plural={}}
\DeclareAcronym{LED}{short = LED, long = light-emitting diode, foreign-plural={}}
\DeclareAcronym{NIST}{short = NIST, long = National Institute of Standards and Technology, foreign-plural={}}
\DeclareAcronym{ICCD}{short = ICCD, long = intensified charge-coupled device, foreign-plural={}}
\DeclareAcronym{QE}{short = QE, long = quantum efficiency, foreign-plural={}}
\DeclareAcronym{Hg}{short = Hg, long = Mercury, foreign-plural={}}
\DeclareAcronym{ZPL}{short = ZPL, long = zero-phonon line, foreign-plural={}}
\DeclareAcronym{NeAr}{short = Ne-Ar, long = Neon-Argon, foreign-plural={}}
\DeclareAcronym{UFBTIS}{short = UFBTIS, long = ulta-fast bias temperature instability setup, foreign-plural={}}
\DeclareAcronym{DAP}{short = DAP, long = donor-acceptor pair, foreign-plural={}}
\DeclareAcronym{LVM}{short = LVM, long = local vibrational mode, foreign-plural={}}
\DeclareAcronym{DLTS}{short = DLTS, long = deep-level transient spectroscopy, foreign-plural={}}
\DeclareAcronym{FWHM}{short = $\FWHM$, long = full width at half maximum, foreign-plural={}}
\DeclareAcronym{CTL}{short = CTL, long = charge transition level, foreign-plural={}}
\DeclareAcronym{HRF}{short = HR-factor, long = Huang-Rhys factor, foreign-plural={}}
\DeclareAcronym{TCAk}{short = (C$_3$)$_\mathrm{Si,k}$, long = cubic tri-carbon antisite, foreign-plural={}}
\DeclareAcronym{TCAh}{short = (C$_3$)$_\mathrm{Si,h}$, long = hexagonal tri-carbon antisite, foreign-plural={}}
\DeclareAcronym{TCIk}{short = (C$_\mathrm{BC}$)$_\mathrm{4,kkkk}$, long = tetra carbon interstitial, foreign-plural={}}
\DeclareAcronym{TCIh}{short = (C$_\mathrm{BC}$)$_\mathrm{4,hhhh}$, long = tetra carbon interstitial, foreign-plural={}}
\DeclareAcronym{DC}{short = DC, long = direct-current, foreign-plural={}}
\DeclareAcronym{CV}{short = CV, long = capacitance-voltage, foreign-plural={}}
\DeclareAcronym{DFT}{short = DFT, long = density-functional theory, foreign-plural={}}

\begin{document}

\preprint{APS/123-QED}

\title{Time-Gated Optical Spectroscopy of Field-Effect Stimulated Recombination via Interfacial Point Defects in Fully-Processed Silicon Carbide Power MOSFETs}

\author{Maximilian W. Feil}
\affiliation{Institute for Microelectronics, TU Wien, Gußhausstraße 27-29 / E360, 1040 Wien, Austria}
\affiliation{Infineon Technologies AG, Am Campeon 1-15, 85579 Neubiberg}
\author{Magdalena Weger}
\affiliation{KAI GmbH, Europastraße 8, 9524 Villach, Austria}
\author{Hans Reisinger}
\affiliation{Infineon Technologies AG, Am Campeon 1-15, 85579 Neubiberg}
\author{Thomas Aichinger}
\affiliation{Infineon Technologies Austria AG, Siemensstraße 2, 9500 Villach, Austria}
\author{André Kabakow}
\affiliation{Infineon Technologies AG, Am Campeon 1-15, 85579 Neubiberg}
\author{Dominic Waldhör}
\affiliation{Institute for Microelectronics, TU Wien, Gußhausstraße 27-29 / E360, 1040 Wien, Austria}
\author{Andreas C. Jakowetz}
\affiliation{Infineon Technologies AG, Am Campeon 1-15, 85579 Neubiberg}
\author{Sven Prigann}
\affiliation{Infineon Technologies AG, Am Campeon 1-15, 85579 Neubiberg}
\author{Gregor Pobegen}
\affiliation{KAI GmbH, Europastraße 8, 9524 Villach, Austria}
\author{Wolfgang Gustin}
\affiliation{Infineon Technologies AG, Am Campeon 1-15, 85579 Neubiberg}
\author{Michael Waltl}
\affiliation{Christian Doppler Laboratory for Single-Defect Spectroscopy at the Institute for Microelectronics, TU Wien, Gußhausstraße 27-29 / E360, 1040 Wien, Austria}
\author{Michel Bockstedte}
\affiliation{Johannes Kepler University Linz, Altenberger Straße 69, 4040 Linz, Austria}
\author{Tibor Grasser}
\affiliation{Institute for Microelectronics, TU Wien, Gußhausstraße 27-29 / E360, 1040 Wien, Austria}

\date{\today}

\begin{abstract}
Fully-processed \acs{SiC} power \acp{MOSFET} emit light during switching of the gate terminal, while both drain and source terminals are grounded. The emitted photons are caused by defect-assisted recombination of electrons and holes at the 4H-\acs{SiC}/\ch{SiO2} interface and can be detected through the \ac{SiC} substrate. Here, we present time-gated spectroscopic characterization of these interfacial point defects. Unlike in previous studies, the devices were opened in such a way that the drain-contact remained electrically active. A separate examination of the photons emitted at the rising and falling transitions of the \acl{VGS} enabled the extraction of two different spectral components. One of these components consists of a single transition with phonon replicas of a \ac{LVM} with an astonishingly high energy of \SI{220}{\milli\electronvolt} -- well above the highest phonon modes in 4H-\acs{SiC} and \acs{SiO2} of \SI{120}{\milli\electronvolt} and \SI{137}{\milli\electronvolt}, respectively. Based on a quantum mechanical model, we successfully fitted its emission spectrum and assigned it to \ac{DAP} recombination involving a carbon cluster-like defect. Other transitions were assigned to $\EHSixSeven$-assisted, $\EKTwo$-D, and nitrogen-aluminum \acl{DAP} recombination. Due to the relevance of these defects in the operation of \ac{SiC} \acp{MOSFET}, these novel insights will contribute to improved reliability and performance of these devices.
\end{abstract}
\acresetall

\maketitle

\section{Introduction}

Wide-bandgap semiconductors, such as \ac{SiC}, \ac{GaN}, and diamond have rapidly moved into the focus of both academic research and industrial development over the past years~\cite{Roccaforte2014,Millan2014,Geis2018}. They are becoming increasingly important and revolutionize power electronics by enabling reliable power conversion with higher energy efficiency at reduced weight and size compared to conventional \ac{Si}-based technologies~\cite{Baliga1989,Zhang2019,Hull2014}. In particular 4H-\ac{SiC} has proven to be the premier wide-bandgap solution at high power and voltages beyond \SI{600}{\volt}, which is required by many renewable-energy applications and emission-free vehicles~\cite{Eddy2009,Langpoklakpam2022}.

However, 4H-\ac{SiC} \acp{MOSFET} exhibit roughly a hundred-times higher defect density at the interface between 4H-\ac{SiC} and its native oxide \ch{SiO2}, compared to the well-studied Si/\ch{SiO2} interface~\cite{Rescher2016a}. Due to the presence of carbon and the use of nitrogen-based annealing techniques~\cite{Li1997,Chung2001}, a variety of new defects can occur compared to \ac{Si}-based devices~\cite{Deak2007a,Devynck2011,Devynck2011a}. These defects can act as traps for charge carriers from both the valence and conduction band of \ac{SiC}. This trapping and detrapping is highly dynamic with time constants down to nanoseconds and lower~\cite{Puschkarsky2018a,Puschkarsky2017,Schleich2021}. The time constants are determined by electric field-dependent activation energies for charge capture and emission, as given by the \ac{NMP} theory~\cite{Huang1950,Kirton1989,Goes2018}. These fast trapping processes do not lead to long-term drifts of \ac{Vth}, but to a hysteresis in all device characteristics such as the transfer or the capacitance voltage characteristics~\cite{Vasilev2023}. Besides this, defect-assisted recombination and subsequently enhanced defect reactions were recently held responsible for a degradation mechanism referred to as gate switching instability~\cite{Aichinger2022,Feil2022a,Feil2023a}. Identifying defects that are involved in the hysteresis and recombination processes at the 4H-\ac{SiC}/\ch{SiO2} interface is therefore elementary for improving the reliability and performance of these devices. For this purpose, a variety of different techniques to characterize and investigate these defects in \ac{SiC} devices has been developed, ranging from purely electrical (measure-stress-measure, capacitance-voltage, deep level transient spectroscopy, charge pumping), over optical (photoluminescence, single-photon-spectroscopy) to magnetic resonance techniques -- either detected electrically or optically~\cite{Grasser2008,Bathen2022}. 

Besides these techniques, Stahlbush \etal~\cite{Stahlbush2001a,Stahlbush2001} and Macfarlane \etal~\cite{Macfarlane2000} observed that 4H-\ac{SiC} \acp{MOSFET} emit light during switching of the gate terminal, while keeping both drain and source terminals grounded. Hereby, the field effect stimulated radiative recombination of electrons and holes via defects at the \ac{SiC}/\ch{SiO2} interface. We have recently confirmed the relation between this light emission and the defects of interest in device operation by correlating the transient \ac{Vth} shift with the light emission~\cite{Feil2023,Feil2022}. This has been investigated further by Weger \etal~\cite{Weger2023,Weger2023a} demonstrating a correlation between the recombination current of a charge pumping experiment and the intensity of a part of the emission spectrum. However, a clear identification of the involved defects was not possible. 

\begin{figure}[t]
\includegraphics[width = \columnwidth]{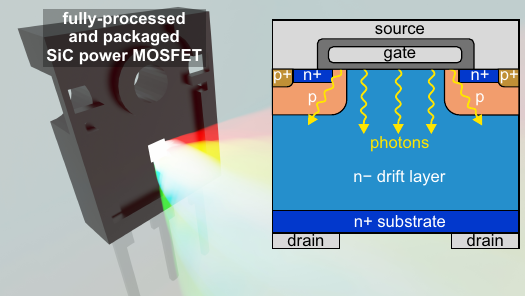}
\caption{\label{fig:IntroOverview} \textbf{Light emission from a \acs{SiC} power \acs{MOSFET} switched between inversion and accumulation.} The light is created via defect-assisted recombination of electrons and holes at the \ch{SiC}/\ch{SiO2} interface. Fully-processed and packaged \ch{SiC} power \acsp{MOSFET} can be opened and prepared for optical inspection, which includes a partial removal of the drain metallization.}
\end{figure}

\begin{figure*}[t]
\includegraphics[width = \textwidth]{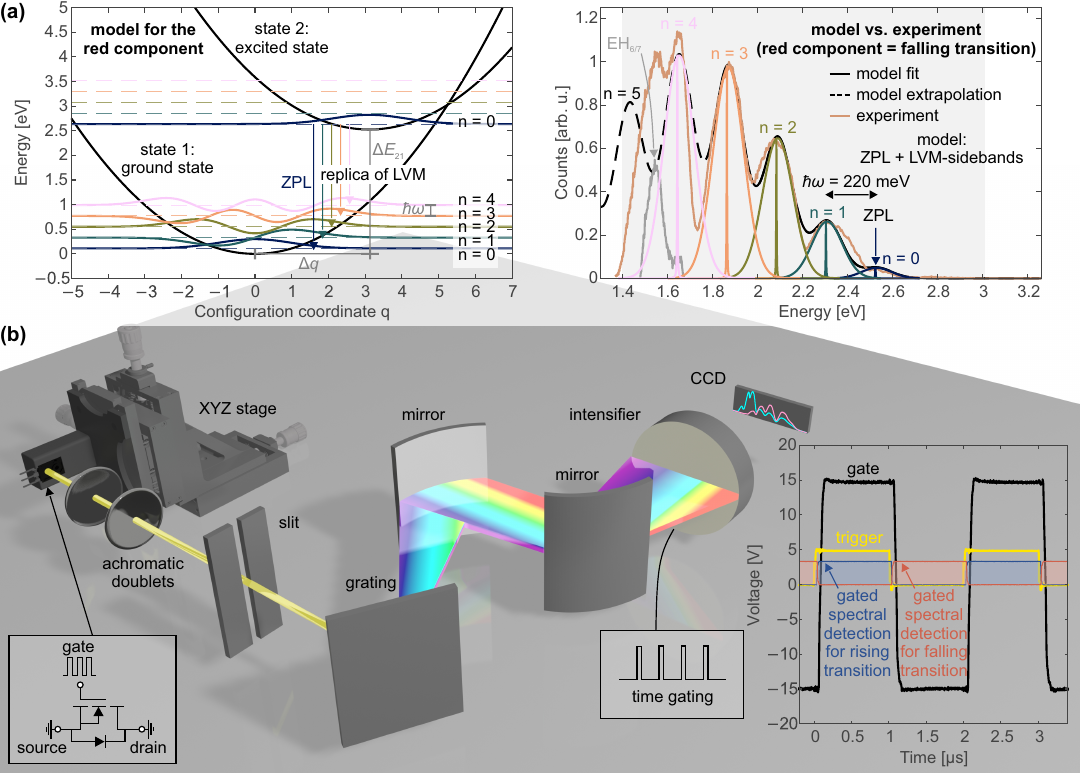}
\caption{\label{fig:ExperimentalSetup} \textbf{Experimental setup and comparison between model and experiment of the emission spectrum of the falling transition (red component).} \textbf{(a)}~The model for the red component consists of a harmonic approximation of the Born-Oppenheimer potentials of two electronic states, between which radiative transitions can emanate from the vibrational ground state of state $2$ to the lowest $5$ vibrational states of state $1$. This includes the \acf{ZPL} and four \acf{LVM} sidebands. The parameters of the potentials can be obtained by fitting the model to an emission spectrum. The right plot shows a comparison between the theoretical emission spectrum from the model in the left plot (obtained by fitting) and the experimentally observed emission spectrum at the falling \ac{VGS}-transition (red component). The grey background indicates the trusted energy region, where the detected spectrum is fully calibrated with respect to intensity and wavelength (see Appendix~\ref{app:Calibration}). \textbf{(c)}~Experimental setup for time-gated optical spectroscopy. The inset on the lower right exemplarily shows the timing of a typical measurement that distinguishes between light emission from the falling and rising transition -- the corresponding spectra are presented in Fig.~\ref{fig:RisingFallingDiode}.}
\end{figure*}

Here, we contribute to the identification by setting up an experiment that allows time-gated spectral detection of the emitted photons during gate switching of a fully processed 4H-\ac{SiC} power \ac{MOSFET} (see Fig.~\ref{fig:IntroOverview}). The time gating allowed us to distinguish between the recombination events occurring at the rising and falling transitions of the \ac{VGS}. Depending on whether the rising or falling transition is being considered, the underlying mechanisms differ considerably, which makes their separation important. At the interface, the distinction between the two transitions arises from the reverse chronological sequence of electron (inversion) and hole (accumulation) presence. This lead to the extraction of two major spectral components. One of them consists of a transition with a \ac{ZPL} of \SI{2.53}{\electronvolt} and phonon replicas with a spacing of \SI{220}{\milli\electronvolt}. Based on a simple model of a one-dimensional harmonic approximation of the involved Born-Oppenheimer potentials, inhomogeneous broadening, and transition rates determined by Fermi's Golden rule (see Fig.~\ref{fig:ExperimentalSetup}a), we identified this component with \ac{DAP} recombination involving a carbon cluster-like defect. This defect is responsible for the observed phonon replicas that stem from a high-energy \ac{LVM}, which is considerably larger than the highest phonon modes in 4H-\acs{SiC} and \ac{SiO2} of \SI{120}{\milli\electronvolt} and \SI{137}{\milli\electronvolt}, respectively. This makes the \ac{LVM} a unique signature of this defect. Besides that, we could further assign an emission peak at \SI{1.55}{\electronvolt} to the $\EHSixSeven$ defect and the other component at the rising transition to other \ac{DAP} recombination paths, including $\EKTwo$-D and nitrogen-aluminum \ac{DAP} recombination. This other component occurring at the rising transition could also be stimulated by forward-biasing the body diode.

First, we describe the employed experimental methodology in Section~\ref{sec:Methods}, followed by a comparison of the emission spectra occurring at the rising and falling transitions, including their relation to the emission spectrum of the forward-biased body diode in Section~\ref{sec:RisingFalling}. Later in Section~\ref{sec:VoltageLevels}, we investigate the dependence of the emission spectrum on the bias levels and in Section \ref{sec:FreqDutyCycle} on frequency and duty cycle. We show the dependence on the transition times of \ac{VGS} in Section~\ref{sec:TransitionTimes}, followed by the temporal evolution of the emission spectrum over the entire \ac{VGS} period in Section~\ref{sec:TemporalResol}. Finally, in Section~\ref{sec:Discussion} we discuss the experimental results in the context of recent experimental and theoretical studies from literature.

\section{Methods}\label{sec:Methods}
We used commercially available, fully processed, n-channel 4H-\ac{SiC} power \acp{MOSFET} with a planar \ac{DMOS} design and in a \ac{TO} package. In these devices, the 4H-\ac{SiC}/\ch{SiO2} interface is located at the (0001) face of the 4H-\ac{SiC} crystal. The three pins of the \ac{TO} package are used to contact the gate, drain, and source terminals. The drain terminal is located at the backside of the chip. We opened the devices from the back for direct optical detection through the \ac{SiC} substrate and epitaxial layer to enable the photons created at the 4H-\ac{SiC}/\ch{SiO2} interface to leave the device without being disturbed~\cite{Weingaertner2002}. We removed the copper lead frame using a wet chemical process with nitric acid and etched away the solder on the chip using aqua regia. Finally, we polished off the backside metallization with diamond paste. Unlike previous studies, this preparation was performed such that some of the drain metallization remained functional to provide electrical control over the potential of the highly doped n-type region at the drain contact.

The \ac{MOSFET} was subsequently mounted on a 3D-printed holder which was in turn mounted on an XYZ translation stage with differential adjusters (see Fig.~\ref{fig:ExperimentalSetup}b). The package was fixed with a screw and its pins were connected via short cables to a custom ultra-fast bias temperature instability setup~\cite{Puschkarsky2018a,Reisinger2006}. All the experiments were conducted at room temperature. For the experiments presented in Section~\ref{sec:TransitionTimes}, where we separately varied the rise and fall time of the \ac{VGS} waveform, we used an Agilent 4156C precision semiconductor parameter analyzer with a 41501B pulse generator and expander unit.

The emitted light was coupled via two achromatic doublet lenses (diameter: \SI{2.54}{\centi\metre}, focal lengths: \SI{30}{\milli\metre}, \SI{100}{\milli\metre}) into the slit of a Teledyne Princeton Instruments IsoPlane 160 imaging spectrograph (Schmidt-Czerny-Turner design) with a grating of \SI{150}{\per\milli\metre} and a \SI{500}{\nano\metre} blaze wavelength combined with silver coated mirrors (see Fig.~\ref{fig:ExperimentalSetup}b). The doublet lenses featured an anti-reflection coating for wavelengths ranging from \SI[print-unity-mantissa=false]{400}{\nano\metre} to \SI[print-unity-mantissa=false]{1100}{\nano\metre}. For time gating the spectral detection, we used a Teledyne Princeton Instruments PI-MAX4 (PM4-256f-HR-FG-18-P43) \ac{ICCD} camera that was attached to the spectrograph.

The benefit of using an \ac{ICCD} camera is its intensifier that can be time-gated in such a way that the \ac{CCD} is exposed to the light only within the \ac{tgate} (min.~\SI{3}{\nano\second}). The time window starts after a defined \ac{tgatedelay} (min.~\SI{25}{\nano\second}) following the transition of a trigger pulse. \SI{50}{\nano\second} after the trigger signal, the ultra-fast bias temperature instability setup switches the gate either from \ac{VL} to \ac{VH} or the other way around (see inset in Fig.~\ref{fig:ExperimentalSetup}b).

This setup allowed time-gated spectral detection in the range of \SIrange[print-unity-mantissa=false]{400}{900}{\nano\metre} with a single exposure of the \ac{CCD}. In all experiments, the amplification of the intensifier was set to unity. We performed both wavelength and intensity calibration, which is described in detail in Appendix~\ref{app:Calibration}. Finally, all recorded spectra were transformed from wavelength into energy space using a Jacobian transformation~\cite{Mooney2013}.

\section{Rising versus Falling Transition}\label{sec:RisingFalling}

The light emission occurs at the transitions of \ac{VGS}~\cite{Feil2022,Feil2023}. Time gating the spectral detection using an \ac{ICCD} camera provides a time resolution that allows the detection of the light emission from the rising or  falling \ac{VGS} transition only. 

\begin{figure}[t]
\includegraphics[width = \columnwidth]{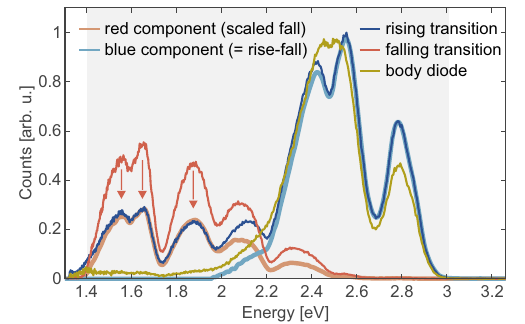}
\caption{\label{fig:RisingFallingDiode} \textbf{Emission spectra from rising and falling \acs{VGS}-transitions and from the forward-biased body diode.} The emission spectrum of the rising transition (rise) is compared to the spectrum of the falling transition (fall). Scaling down the emission of the falling transition shows good agreement with the spectrum of the rising transition and is referred to as the red component. The emission besides the red component at the rising transition is called the blue component and agrees with the emission of the body diode.}
\end{figure}

\begin{figure}[t]
\includegraphics[width = \columnwidth]{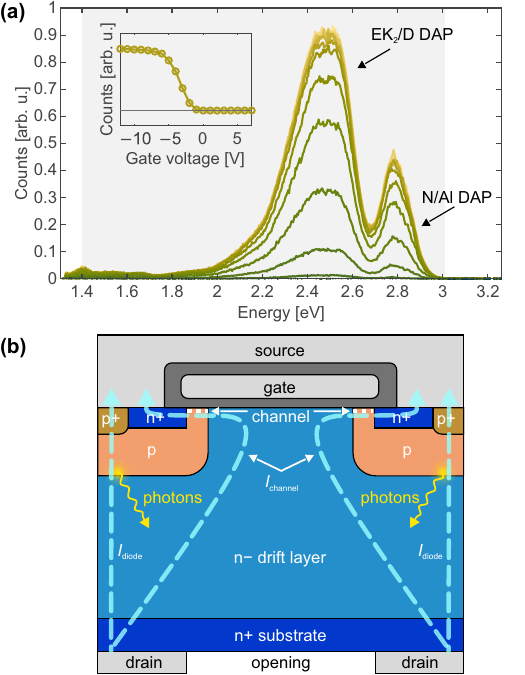}
\caption{\label{fig:BodyDiode} \textbf{Light emission from the forward-biased body diode.} \textbf{(a)}~Emission spectra obtained by applying a forward bias to the body diode at a constant current but varying \acs{VGS}. The inset shows the dependence of the integrated photon counts on \acs{VGS}. \textbf{(b)}~Schematic illustration of the \acs{MOSFET} cell and the currents flowing upon a forward biased body diode.}
\end{figure}

To separate the emission spectra of the rising and falling transitions, we used continuous gate switching at \SI{500}{\kilo\hertz} between $\VL = \SI{-15}{\volt}$ and $\VH = \SI{15}{\volt}$ with \SI{50}{\percent} duty cycle and \ac{trise} and \ac{tfall} of \SI{50}{\nano\second}. We chose $\tgatedelay = \SI{25}{\nano\second}$ and $\tgate = \SI{1}{\micro\second}$ to fully capture the light emission of the respective \ac{VGS}-transition, while rejecting the light emission of the other \ac{VGS} transition. In a separate experiment, we investigated the light emission that occurs when a forward bias is applied to the body diode. As the underlying recombination events occur dominantly in the bulk of the 4H-\ac{SiC} crystal, we can use the resulting spectrum to compare the dominant recombination paths in the bulk to the events found closer to the 4H-\ac{SiC}/\ch{SiO2} interface that are triggered by switching the gate terminal. We measured the spectra from the body diode with a forward bias corresponding to a constant current of \SI{40}{\milli\ampere}.

\begin{figure*}[t]
\includegraphics[width = \textwidth]{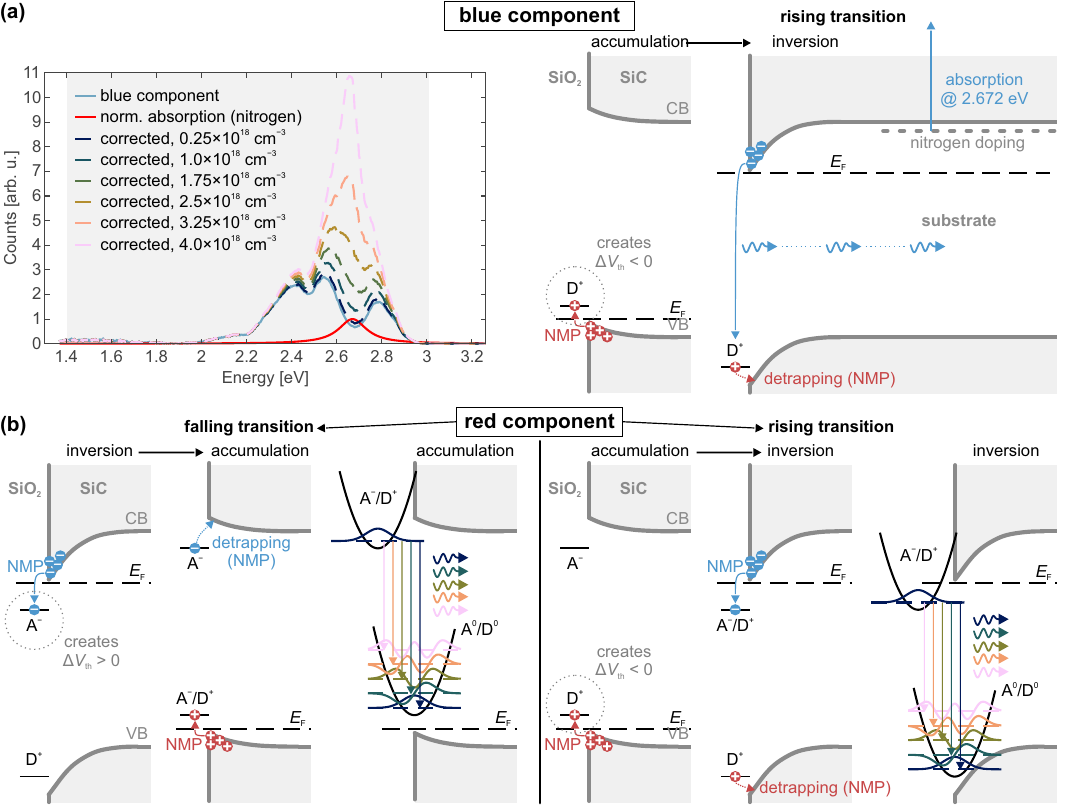}
\caption{\label{fig:BandDiagram} \textbf{Radiative transitions via interfacial point defects that constitute the red and blue components.} \textbf{(a)}~A portion of the blue component is probably absorbed by a transition from the nitrogen doping to a higher state in the conduction band.  The impact of the absorption was calculated and plotted for different doping concentrations. According to literature, the blue component can mostly be assigned to \ac{DAP} recombination between a D-center and an EK$_2$ center and nitrogen and aluminum dopants. However, in the here presented experiments, the blue component behaves more like a donor-like defect close to the valence band. \textbf{(b)}~The red component occurs both at the rising and falling \ac{VGS}-transitions and consists of multiple transitions from the vibrational ground state of the electronic state $2$ (A$^{-}$/D$^{+}$) to the lowest five vibrational states of the electronic state $1$ (A$^{0}$/D$^{0}$). Prior to recombination, both an electron and a hole need to be trapped in the respective defect state.}
\end{figure*}

The spectra obtained on switching the gate terminal are shown in Fig.~\ref{fig:RisingFallingDiode}. First, it is notable that the spectrum at the rising transition covers a broader range of energy up to about \SI{3}{\electronvolt}, whereas the spectrum at the falling transition disappears for energies above \SI{2.65}{\electronvolt}. We found good agreement between the two emission spectra for energies in the range of \SIrange{1.6}{2.0}{\electronvolt} by scaling down the emission spectrum of the falling transition such that it matched the emission spectrum of the rising transition. Most importantly, the difference between the scaled spectrum of the falling transition and the spectrum of the rising transition results in a spectrum that approximately matches the emission of the forward-biased body diode. The emission spectra of the body diode for a constant current, but varying \ac{VGS}, are shown in Fig.~\ref{fig:BodyDiode}a. Aside from the overall intensity, the spectrum does not depend on \ac{VGS} -- we coherently obtained the same spectrum for each \ac{VGS} condition and consequently the same bulk recombination pathways at the body diode. The variation in the overall intensity can be explained by the variation in current through the diode that decreases with increasing \ac{VGS} and the corresponding rise in current through the channel (see Fig.~\ref{fig:BodyDiode}b). This effect has previously been used for condition monitoring~\cite{Susinni2020,Ruppert2022}.

The finding that the emission spectrum at the rising transition is rather blueish compared to the emission at the falling transition agrees with a previous study~\cite{Macfarlane2000}. One of the most important new insights was that the emission spectrum of the falling transition is a subset of the emission spectrum of the rising transition. The spectra only differ by a scaling factor (see Fig.~\ref{fig:RisingFallingDiode}). Consequently, the recombination process of the falling transition equally occurs at the rising transition. Finally, the additionally observed emission at the rising transition is similarly present under a forward-bias condition of the body diode.

Based on their spectral location and associated human-perceived color (red: \SIrange{1.6}{2.0}{\electronvolt}, blue: \SIrange{2.5}{2.8}{\electronvolt}), we refer to the common spectral component of rising and falling transitions as the ``red component" and to the additional spectral component of the rising transition as the ``blue component". Hereafter, we started a separate investigation of their behavior by modeling each spectrum as a superposition of these components (see Fig.~\ref{fig:RisingFallingDiode}).

\subsection{The Blue Component}\label{sec:BlueComp}

The fact that the additional emission spectrum occurring at the rising transition matches the emission spectrum of the body diode suggests that this blue component can be assigned to the same or similar types of 4H-\ac{SiC} bulk defects, even though these emitting defects must be located at or close to the 4H-\ac{SiC}/\ch{SiO2} interface. As 4H-\ac{SiC} bulk defects, including those occurring at pn-junctions, have been thoroughly investigated in the past, we can tentatively assign the peak around \SI{2.5}{\electronvolt} to \ac{DAP} recombination between the so-called D-center and a donor defect EK$_2$, which are both known from \ac{DLTS}~\cite{Kuznetsov1995,Klein2006} with possibly further spectral contributions from $\ZOneTwo$ centers~\cite{Fabbri2009}. Furthermore, the peak around \SI{2.8}{\electronvolt} is assigned to \ac{DAP} recombination between nitrogen and aluminum dopants~\cite{Bishop2009,Kuznetsov1995}. Defect-assisted recombination involving the $\DOne$ center could also contribute to this peak~\cite{Fissel2001,Weger2023}.

Although the blue component could potentially be assigned to \ac{DAP} recombination based on the mentioned literature~\cite{Kuznetsov1995,Klein2006,Fabbri2009,Bishop2009,Kuznetsov1995,Fissel2001,Weger2023}, our experimental results hint towards a donor-like defect close to the valence band and subsequent recombination with channel electrons (see Fig.~\ref{fig:BandDiagram}a). The discrepancy between literature and our results could be related to extremely short trapping/detrapping time constants of the second defect state close to the conduction band, which would let the recombination appear to occur in direct interaction with the conduction band. This would explain why the blue component only appears at the rising transition and its dependence on the transition time (see Section~\ref{sec:TransitionTimes}).

Note that an inconspicuous portion of the blue component might be absorbed by the 4H-\ac{SiC} substrate. Based on absorption measurements from a transition from nitrogen doping to a higher state in a previous study~\cite{Weingaertner2002}, we assume a Lorentzian-shaped absorption with a peak absorption coefficient of $\alpha = \alpha_0 + \left( \kappa \cdot c_\mathrm{n} \right)$, where $\kappa = \SI{3.6E-17}{\centi\metre\squared}$, $\alpha_0 = \SI{2.4}{\per\centi\metre}$, and $c_\mathrm{n}$ is the n-type nitrogen doping concentration. Fig.~\ref{fig:BandDiagram}a shows the blue component, the normalized absorption peak, and the blue component corrected by the absorption of a substrate of \SI{185}{\micro\metre} for several typical doping concentrations. It should be noted that the absorption peak coincides well with a valley in the emission spectrum.

\subsection{The Red Component}\label{sec:RedComp}

Interestingly, most of the emission peaks of the red component are equally spaced in energy, which reminds of phonon replicas of a single transition. As the energy spacing is with about \SI{220}{\milli\electronvolt} -- far higher than the highest phonon modes in 4H-\ac{SiC} or \ac{SiO2} of \SI{120}{\milli\electronvolt} and \SI{137}{\milli\electronvolt}, respectively -- these replicas must be related to a \ac{LVM} that can be modeled by a quantum mechanical harmonic oscillator~\cite{McCluskey2000}. The potential energy surface of this harmonic oscillator is conventionally obtained from a one-dimensional harmonic approximation of the involved Born-Oppenheimer potentials. Introducing the potentials of two hypothetical states, $1$ (ground state) and $2$ (excited state), that are illustrated in Fig.~\ref{fig:ExperimentalSetup}a, allowed us to consistently fit the emission spectrum of the red component. As discussed in detail in Appendix \ref{app:QMModel}, the emission spectrum, $I \left( \Ephoton \right)$, of such a system, including broadening of transition energies, is given by
\begin{equation}
I \left( \Ephoton \right) \propto \sum_{m,n} \omegaphoton^3 \cdot \abs{\bra{\phi_{m}^{2}}\ket{\phi_{n}^{1}}}^2 \cdot L \left(\Ephoton, E_{mn}, \sigma_{mn} \right).
\label{eq:ModelEmissionSpectrum}
\end{equation}
Here, $\omegaphoton$ is the angular frequency of the emitted photon, $\bra{\phi_{m}^{2}}\ket{\phi_{n}^{1}}$ are the transition matrix elements, $L$ is the lineshape function of the respective transition, $\Ephoton$ is the energy of the emitted photon, and $E_{mn}$ and $\sigma_{mn}$ are the transition energy and its standard deviation, respectively.

Consequently, $I \left( \Ephoton \right)$ is determined by four parameters ($\dETwoOne$, $\hbar \omega_1$, $\hbar \omega_2$, $\Delta q$) describing the two harmonic potentials and four parameters ($\sigma_{1}$, $\sigma_{2}$, $\sigma_{\Delta E}$, $\dELor$) describing the broadening effects of the emission peaks. Typically, it is assumed that $\omega := \omega_1 = \omega_2$ and $\sigma := \sigma_{1} = \sigma_{2}$, which reduced the total number of parameters from eight to six.

As shown in Fig.~\ref{fig:ExperimentalSetup}a, fitting this model to the emission spectrum of the falling transition for energies above $\SI{1.77}{\electronvolt}$ produces an excellent agreement. The parameters and their values are listed in Table~\ref{tab:DefectParameters}. We obtained a \ac{ZPL} of \SI{2.53}{\electronvolt} and the emission lines were separated by $\hbar \omega = \SI{220}{\milli\electronvolt}$. The standard deviation $\sigma$ of $\hbar\omega$ and the \ac{FWHM} of the Lorentz lineshape $\dELor$ were very small or even negligible. Consequently, broadening of the emission peaks is predominantly caused by the distribution of the \acp{CTL}, resulting in a Gaussian distributed $\dETwoOne$ with a $\FWHM = 2 \sqrt{2 \ln(2)} \sigma_{\Delta E} = \SI{158}{\milli\electronvolt}$. Assuming that the standard deviations of the two \acp{CTL} are equal, their \ac{FWHM} would be \SI{112}{\milli\electronvolt}. 

\begin{table}[h]
\centering
\caption{\label{tab:DefectParameters}
Defect parameters obtained from the fit shown in Fig.~\ref{fig:ExperimentalSetup}a.}
\begin{tabular}{c|cccccc}
\multirow{2}{*}{Parameter} & $\dETwoOne$ & $\hbar \omega$ & $\Delta q$ & $\sigma$ & $\sigma_{\Delta E}$ & $\dELor$ \\
 & [\SI{}{\electronvolt}] & [\SI{}{\milli\electronvolt}] &  & [\SI{}{\milli\electronvolt}] & [\SI{}{\milli\electronvolt}] & [\SI{}{\milli\electronvolt}] \\
\hline
Value & \SI{2.53}{} & \SI{220}{} & \SI{3.11}{} & \SI{0}{} & \SI{67}{} & \SI{2.1}{}
\end{tabular}
\end{table}

Remarkably, the model nicely predicts the fifth emission peak around \SI{1.65}{\electronvolt}, but fails with the peak around \SI{1.55}{\electronvolt}. This additional emission peak is indicated in Fig.~\ref{fig:ExperimentalSetup}a as the difference between measured emission spectrum and the extrapolated model. We assigned this additional peak to another type of defect, which is probably the $\EHSixSeven$ center~\cite{Weger2023}.

\begin{figure*}[t!]
\includegraphics[width = \textwidth]{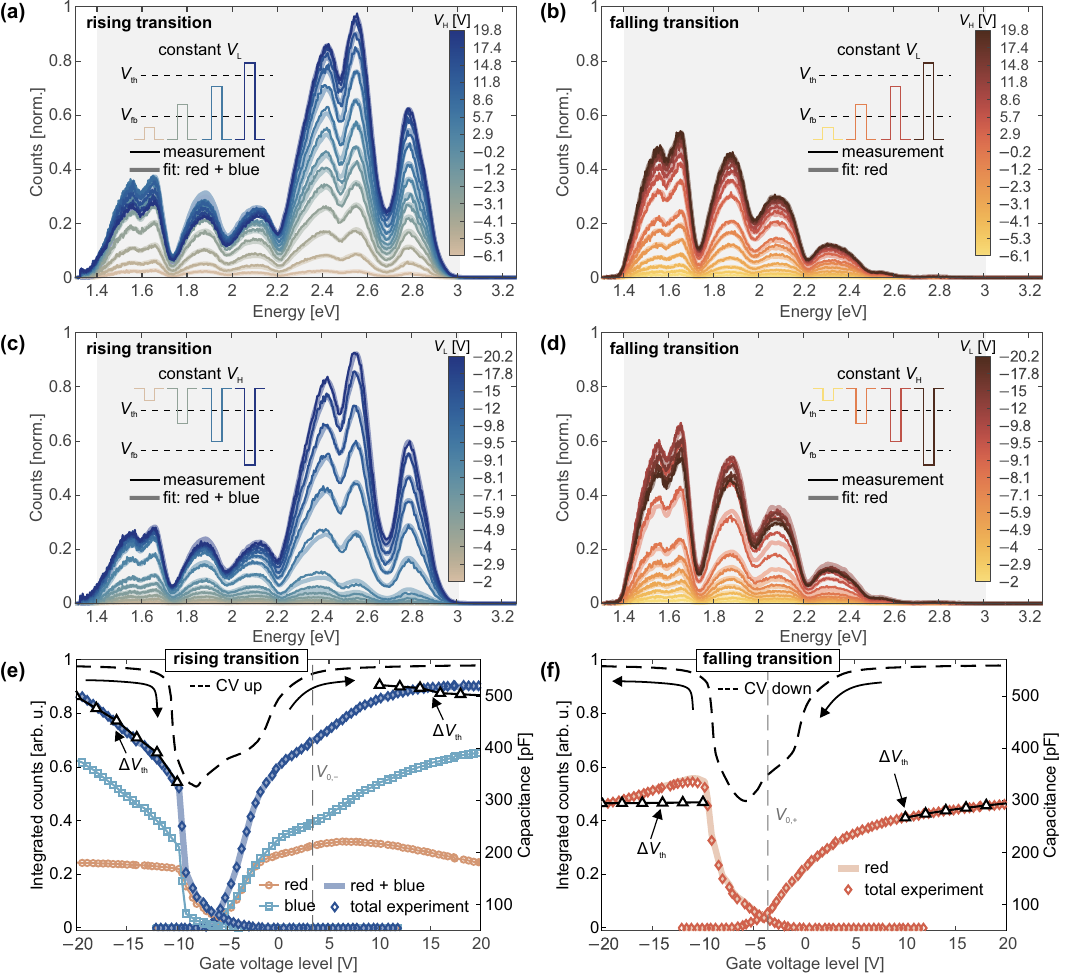}
\caption{\label{fig:GateVoltageSweeps} \textbf{Bias level dependence.} \textbf{a--b} Emission spectra of the rising and falling transitions, respectively, while keeping \ac{VL} constant. \textbf{c--d} Emission spectra of the rising and falling transitions, respectively, while keeping \ac{VH} constant. \textbf{e--f} Integrated photon counts of the emission spectra at the rising and falling transitions, respectively, including constant \ac{VL} and \ac{VH} curves and their decomposition into the red and blue components. The corresponding up and down sweeps of the \ac{CV} characteristic and the fits based on equation~(\ref{eq:ICVth}) are shown as well.}
\end{figure*}

Unfortunately, the model does not provide information on the absolute position of states $1$ and $2$ within the bandgap of 4H-\ac{SiC} ($\Egap = \SI{3.26}{\electronvolt}$). If the two states were centered within the bandgap, they would be $ \left( \Egap - \dETwoOne \right)/2 = \SI{0.36}{\electronvolt}$ away from the band edges. The fact that the red component appears both at the rising and the falling transition leads to the conclusion that it must be related to \ac{DAP} recombination. In the context of the red component, we defined a donor D as a defect that emits an electron irrespective of its charge state and an acceptor A as a defect that captures an electron irrespective of its charge state. This is illustrated in Fig.~\ref{fig:BandDiagram}b. Independent of the type of transition, the red component is created by a transition from state $2$ (A$^{-}$/D$^{+}$) to $1$ (A$^{0}$/D$^{0}$). Prior to this transition, the other transitions between valence and conduction bands and the states A$^{-}$ (falling transition) and D$^{+}$ (rising transition) are \ac{NMP} transitions, that are strongly bias dependent and allow the linking of the transient shift of \ac{Vth} to the photon emission~\cite{Feil2023}. This can be observed in the dependence of the photon emission on the bias levels, which is investigated in the next section.

\section{Bias Levels}\label{sec:VoltageLevels}

\begin{figure*}[t]
\includegraphics[width = \textwidth]{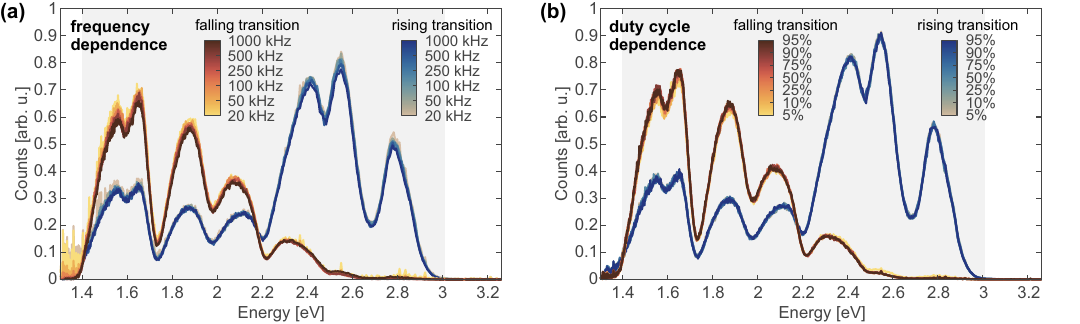}
\caption{\label{fig:DifferentFrequencies} \textbf{Frequency and duty cycle independence.} \textbf{(a)}~The frequency dependence of the emission spectra at the rising and falling transitions. \textbf{(b)}~The duty cycle dependence of the emission spectra at the rising and falling transition.}
\end{figure*}

\Ac{DAP} recombination via the discussed defect states inherently involves trapping and detrapping of electrons and holes during the application of \ac{VH} or \ac{VL}, respectively. It is therefore expected that the properties of the prominent transient \ac{Vth} shift will be reflected in the properties of the light emission. We have already observed such correlations for both stress and recovery in our previous studies~\cite{Feil2023,Feil2022}. In fact, the setting considered here is similar to charge pumping experiments~\cite{Groeseneken1984,Salinaro2015}, where the measured recombination current is analogous to the measured photon flux. However, it should be noted that every recombination pathway that contributes to a charge pumping current will not necessarily involve a radiative transition that can be detected by optical spectroscopy.

In this section, we investigate how the red and blue spectral components depend on the gate bias levels, \ac{VH} and \ac{VL}. Analogously to the constant \ac{VL}-technique in charge pumping experiments, we kept $\VL = \SI{-20}{\volt}$ constant in deep accumulation, successively increase \ac{VH}, and measured the emission spectrum separately at the rising and falling \ac{VGS}-transition. Besides that, we kept $\VH = \SI{20}{\volt}$ constant in deep inversion and successively decreased the low level, which is the analogue to the constant \ac{VH}-technique in charge pumping experiments. In total, we obtained four sets of spectra that are shown in Figs.~\ref{fig:GateVoltageSweeps}a--d. As mentioned in Section~\ref{sec:RisingFalling}, we can decompose each spectrum at the rising transition and determine the amplitudes of the red and blue components by fitting them to the measured spectrum. Hence, only two fitting parameters representing the amplitudes of the components are needed. Integrating the spectral components yields the associated total number of photon counts, which is shown for both rising and falling transitions in Figs.~\ref{fig:GateVoltageSweeps}e--f.

Along with optical spectroscopy, we utilized impedance spectroscopy to analyze the \ac{CV} characteristic of the tested \ac{MOSFET}. This characteristic is well correlated with the optical emission intensity shown in Figs.~\ref{fig:GateVoltageSweeps}e--f. A significant increase in light intensity was observed for each component when the device entered inversion or accumulation. While the red component at the rising transition showed a rather \ac{VGS}-independent behavior, the blue component increased significantly with either decreasing or increasing bias. For the falling transition, the red component increased with increasing \ac{VH} and saturated or even decreased with decreasing \ac{VL}.

We also performed ultra-fast \ac{Vth} measurements to compare the light emission with the short-term charge trapping behavior. For this purpose, we used a pristine device with intact drain metallization. Then, we measured \ac{Vth} at the end of the \ac{VL}- or \ac{VH}-phase before the respective \ac{VGS} transition after \SI{4.1}{\milli\second} of \SI{500}{\kilo\hertz} switching between \SI{20}{\volt} and \SI{-20}{\volt} with a transition time of \SI{50}{\nano\second}. The measurement delay time was set to \SI{1}{\micro\second}. Following our approach in \cite{Feil2023} for double pulses, we fit the threshold voltage data to the integrated photon counts ($\mathrm{IC}$) as per
\begin{equation}\label{eq:ICVth}
\mathrm{IC} = C_s |\Vth \left( \VGS \right) - V_{0,s}|, \ s \in \{+,- \}.
\end{equation}
Here, $V_{0,s}$ and $C_s$ are fitting parameters for the rising ($+$) and falling ($-$) transitions, respectively. Despite some differences that probably arose from extensive, continuous switching, we found good agreement between the \ac{Vth} measurement and the integrated photon counts.

In summary, the device exhibited light emission only upon switching between inversion and accumulation. This was observed by the strong increase in photon counts around the threshold and flat band voltages that were determined by the \ac{CV} characteristic. Finally, the sum of all components correlated with the \ac{Vth} measured before the respective \ac{VGS} transition. Considering that \ac{Vth} is directly related to the amount of trapped charge and the absence of spectral changes within the components themselves, the observations agree with the mechanisms illustrated in Fig.~\ref{fig:BandDiagram}.

\section{Frequency and Duty Cycle}\label{sec:FreqDutyCycle}

\begin{figure*}[]
\includegraphics[width = \textwidth]{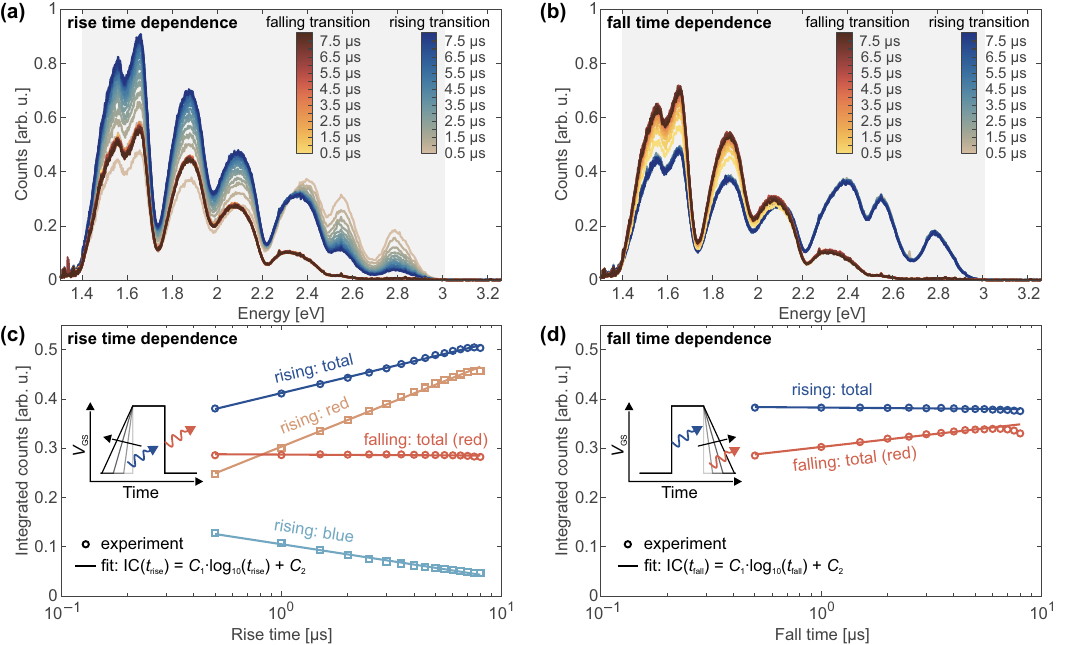}
\caption{\label{fig:TransitionTimes} \textbf{Transition time dependence.} \textbf{(a)}~Rise time dependence of the emission spectra at the rising and falling transitions. \textbf{(b)}~Fall time dependence of the emission spectra at the rising and falling transitions. \textbf{(c)}~Rise time dependence of the integrated photon counts of the total emission spectrum and the red and blue components. \textbf{(d)}~Fall time dependence of the integrated photon counts of the total emission spectrum and the red and blue components.}
\end{figure*}

Besides the bias levels, we also varied the switching frequency in the range of \SIrange{20}{1000}{\kilo\hertz} for a fixed duty cycle of \SI{50}{\percent} and the duty cycle in the range of \SIrange{5}{95}{\percent} for a fixed frequency of \SI{100}{\kilo\hertz}. In these experiments, all spectra consisted of \SI{2E6}{} \ac{VGS} transitions and the bias levels were set to \SI{15}{\volt}/\SI{-15}{\volt}. The results are shown in Fig.~\ref{fig:DifferentFrequencies}. Despite small changes around \SI{1.6}{\electronvolt} in the emission spectrum of the falling transition (see Fig.~\ref{fig:DifferentFrequencies}a), we did not find any significant impact of either frequency or duty cycle. Therefore, the initial trapping of the later recombining charge carriers occurs very fast. At the bias levels used, the corresponding capture time constants needed to be well below \SI{500}{\nano\second}.

\section{Transition Times}\label{sec:TransitionTimes}

In charge pumping experiments, the recombination current typically decreases with increasing transition time. This is caused by detrapping of previously trapped charge carriers during the gate voltage transition from inversion to accumulation or vice versa. Hereby, the charge pumping current typically exhibits a linear relationship to the logarithm of the transition time~\cite{Groeseneken1984}. 

Analogously, we investigated the light emission at the rising and falling transitions separately while varying either the rise or the fall time. For this purpose, we used a \SI{20}{\volt}/\SI{-20}{\volt}, \SI{50}{\kilo\hertz} gate waveform and separately varied the transition times from \SI{0.5}{\micro\second} up to \SI{8}{\micro\second}, while keeping the other transition time constant at \SI{0.5}{\micro\second}. Analogous to the investigation of the bias level dependence, discussed in Section~\ref{sec:VoltageLevels}, we extracted the red and blue components from the emission spectra of the rising transition to determine their respective dependency.

The results are shown in Fig.~\ref{fig:TransitionTimes}. We found for all spectral components the same linear relation to the logarithm of the transition time. First, the emission spectrum at the rising transition was not affected by the fall time and the emission spectrum at the falling transition was not affected by the rise time. Second, although the long-term limit is certainly a decreasing light emission with increasing transition time, we found, in the investigated regime of transition times, an increasing red component with both increasing rise time at the rising transition and increasing fall time at the falling transition. In contrast to that, the blue component at the rising transition decreases with increasing rise time, as typically observed for the recombination current through interface defect states in charge pumping experiments.

The independence of radiative recombination at the falling transition from the rise time and at the rising transition from the fall time agrees with the postulated short capture time constants from Section~\ref{sec:FreqDutyCycle}. The decreasing blue component with the rise time at the rising transition rather agrees with a donor-like state close to the valence band (see Fig.~\ref{fig:BandDiagram}). The increasing red component with the rise time at the rising transition and with the fall time at the falling transition is related to a peculiarity of \ac{DAP} recombination. In contrast to the recombination between a defect state at the interface and either the valence or conduction band, which is the basic assumption in charge pumping experiments, both the donor and the acceptor state have to first capture and sustain a charge carrier so that recombination between them can occur. Consequently, depending on the involved time constants, there is a regime of increasing recombination with increasing transition time. This is discussed in greater detail in Section~\ref{sec:CompDAPDetrapping}.

\section{Temporal Resolution}\label{sec:TemporalResol}

\begin{figure*}[t]
\includegraphics[width = \textwidth]{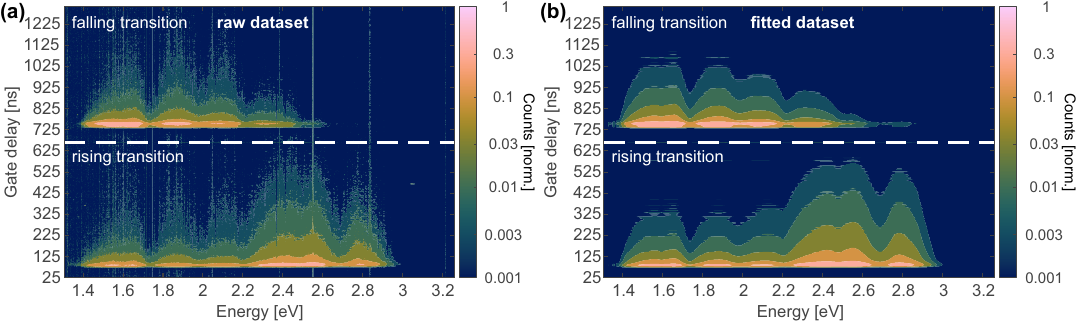}
\caption{\label{fig:TemporalMaps} \textbf{Contour plot of time-resolved emission spectra over an entire gate voltage period of 1280~ns.} Each horizontal line corresponds to an emission spectrum for the respective gate delay indicated on the vertical axis. Each spectrum was measured within a gate time window of \SI{3}{\nano\second}. The gate delay time was successively increased to scan the entire gate voltage period. \textbf{(a)}~Normalized raw dataset. \textbf{(b)}~Normalized dataset fitted with red and blue components.}
\end{figure*}

\begin{figure}[]
\includegraphics[width = \columnwidth]{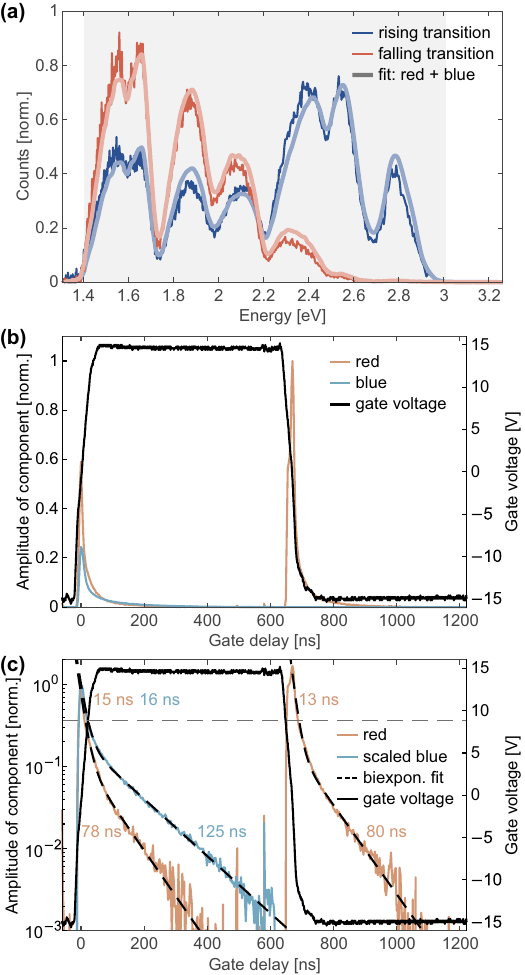}
\caption{\label{fig:TemporalResolution} \textbf{Temporal evolution of the red and blue components over one gate voltage period.} \textbf{(a)}~Exemplary fits of emission spectra from the total dataset shown in Fig.~\ref{fig:TemporalMaps}. \textbf{(b)}~Temporal evolution of the two components on a linear y-scale. \textbf{(c)}~Temporal evolution of the two components on a logarithmic y-scale.}
\end{figure}

As mentioned in Section~\ref{sec:Methods}, the used \ac{ICCD} camera features a mininum \ac{tgate} of \SI{3}{\nano\second}. Apparently, this not only allows the detection of the temporal evolution of the emission spectrum over an entire gate voltage period, but also resolves the temporal evolution during the \ac{VGS} transition itself. 

For this experiment, we used a frequency of \SI{78.1}{\kilo\hertz}, corresponding to a period of \SI{1.28}{\micro\second}, with voltage levels of \SI{15}{\volt}/\SI{-15}{\volt} and a transition time of \SI{50}{\nano\second}. We used the mentioned minimum gate width of \SI{3}{\nano\second}. Again, we extracted red and blue components from each spectrum. Fig.~\ref{fig:TemporalMaps}a shows the raw data of the entire experiment and Fig.~\ref{fig:TemporalMaps}b presents the corresponding fitted data set, consisting of a superposition of red and blue components. Hereby, each spectrum was fitted independently. Using two exemplary spectra, the quality of the fit is shown in Fig.~\ref{fig:TemporalResolution}a. The two components from Fig.~\ref{fig:RisingFallingDiode} nicely reproduced the observed spectra. Subsequently, we investigated the red and blue components separately. The amplitudes of the spectral components are plotted with a linear y-axis in Fig.~\ref{fig:TemporalResolution}b and with a logarithmic y-axis in Fig.~\ref{fig:TemporalResolution}c.

As already found in our previous work \cite{Feil2022,Feil2023}, the light emission is on a linear scale strongly localized at the switching transitions. On the logarithmic scale, shown in Fig.~\ref{fig:TemporalResolution}c, all three decay curves show a biexponential decay. Consequently, the decay curves can be described by
\begin{align}
I \left( \tgatedelay \right) =& C \exp\left( -\frac{\tgatedelay - t_0}{\tau_1} \right) \nonumber \\
&+ \left(1 - C \right) \exp \left( -\frac{\tgatedelay - t_0}{\tau_2} \right).
\label{eq:biexponential}
\end{align}
For each normalized component, $t_0$ is the time of maximum photon emission. The parameter $C$ represents the share of the first exponential decay, and $\tau_1$ and $\tau_2$ are the respective decay time constants. These parameters can be obtained by fitting and are listed in Table~\ref{tab:BiexponentialParameters}. 

Apparently, the red component exhibits the same $\tau$ and $C$ for both transitions. As the decay with $\tau_1$ overlaps in time with the gate voltage transitions themselves, the deduction of the underlying physical mechanism is rather complicated. However, the decay with $\tau_2$ occurs during the phase of constant gate voltage. As the $\tau_2$'s agree, we can conclude that the underlying recombination mechanism must be identical for both transitions. The blue component has about the same short decay time constant $\tau_1$, but differs significantly in $\tau_2$, which is almost \SI{60}{\percent} larger compared to that of the red component.

\begin{table}[]
\centering
\caption{\label{tab:BiexponentialParameters} Parameters obtained by fitting equation~(\ref{eq:biexponential}) to the data in Fig.~\ref{fig:TemporalResolution}c.}
\begin{tabular}{c|ccc|ccc}
 & \multicolumn{3}{c|}{red} & \multicolumn{3}{c}{blue} \\
 & $C$ & $\tau_1$[\SI{}{\nano\second}] & $\tau_2$[\SI{}{\nano\second}] & $C$ & $\tau_1$[\SI{}{\nano\second}] & $\tau_2$[\SI{}{\nano\second}] \\
\hline
rising & 0.87 & 15 & 78 & 0.80 & 16 & 125 \\
falling & 0.89 & 13 & 80 & -- & -- & -- 
\end{tabular}
\end{table}

The observation that the radiative decay time constants are identical for both rising and falling transitions agrees with the proposal made in Section~\ref{sec:RisingFalling} -- the red component stems from \ac{DAP} recombination and associated vibrational sidebands. The reason for the higher $\tau_2$ of the blue component is, as of now, not entirely clear. However, as the red components agree, it confirms our understanding of splitting the total spectrum at the rising transition into the two components.

\section{Discussion}\label{sec:Discussion}

As discussed in the previous sections, the underlying defects belonging to the red and blue components are most probably very different. The idea presented in Section~\ref{sec:RisingFalling}, of decomposing all spectra into their respective components, was useful to study their individual properties. Based on the fitting of the red component arising at the falling transition of \ac{VGS}, we identified $\dETwoOne$, $\hbar \omega$, and $\Delta q$ as physical defect parameters that can be used to search for possible defect candidates that might be the origin of this component.

\subsection{Carbon Clusters as the Cause of the Red Component}

As shown in Section~\ref{sec:RisingFalling}, the observed emission spectra at the falling transition clearly exhibit the signature of radiative recombination between the vibrational ground state of an excited electronic state and vibrational sidebands of an electronic ground state. The spacing $\hbar \omega = \SI{220}{\milli\electronvolt}$ of the vibrational sidebands is considerably higher than the highest optical phonon modes of about \SI{120}{\milli\electronvolt} and \SI{137}{\milli\electronvolt} in 4H-\ac{SiC} and \ch{SiO2}, respectively~\cite{Serrano2002,Serrano2003,Deng2019}. Thus, the observed vibrational sidebands must stem from \acp{LVM} of the underlying defect, which can be higher in energy than collective lattice vibrations~\cite{McCluskey2000}.

Indeed, several different \acp{LVM} with energies of up to \SI{247}{\milli\electronvolt} have been detected in 4H-\ac{SiC} in low temperature photoluminescence measurements~\cite{Steeds2008,Mattausch2006}. These \acp{LVM} are exclusively linked to carbon clusters~\cite{Mattausch2004,Bockstedte2004,Li2023b,Gali2003}, whose emission intensity and \ac{ZPL} depend on quantities such as doping concentration and annealing temperature~\cite{Steeds2004,Steeds2008}. Carbon clusters likely exist in a similar fashion at the 4H-\ac{SiC}/\ch{SiO2} interface. Photoluminescence from interface defects, mostly around \SIrange{1.55}{2.48}{\electronvolt}, not only exhibited \acp{LVM} of up to \SI{220}{\milli\electronvolt}, but could also successfully be linked to the interface-defect density extracted by capacitance-voltage measurements~\cite{Woerle2022,Johnson2019}. Note that the \ac{LVM} and the energy range perfectly agree with the red component. Most probably the same defects were shown to be single-photon emitters, whereby their polarization hints towards defects at the \ac{SiC}-side of the interface~\cite{Abe2018,Lohrmann2015,Lohrmann2016}. It was speculated that the observed single-photon emitters were carbon or oxygen related interface defects, as they appeared after thermal oxidation of the \ac{SiC} surface and disappeared again after the oxide was removed. Two-dimensional mapping of these defects was performed using confocal photoluminescence and tunneling electroluminescence~\cite{Woerle2019,Alyabyeva2022} as well as capacitive methods~\cite{Yamagishi2017}. In all the cases, the observed defects predominantly appeared at the bumps on the (0001) crystal surface.

As mentioned earlier, the fact that the red component, which is suspected to originate from carbon clusters, appears both at the falling and the rising transitions leads to the conclusion that the underlying process requires to trap both electron and hole in acceptor (A$^{-}$) and donor (D$^{+}$) states first, before the actual recombination event occurs (Fig.~\ref{fig:BandDiagram}b). Otherwise, the band electrons or holes would rapidly disappear during the switching event, such that no recombination event would occur. These donor and acceptor states bind the respective charge carrier long enough for the opposite charge carrier to also get trapped, so that they can subsequently recombine.

Comparing our observed \ac{LVM} with theoretical and experimental results from literature for bulk 4H-\ac{SiC}, it might be suspected that the observed \ac{LVM} stems either from cubic and/or hexagonal tri-carbon antisite clusters (\acuse{TCAk}\ac{TCAk} and \acuse{TCAh}\ac{TCAh}) or cubic and/or hexagonal tetra carbon interstitials (\acuse{TCIk}\ac{TCIk} and \acuse{TCIh}\ac{TCIh})~\cite{Li2023b, Mattausch2004,Mattausch2006}, that show similar values for \ac{ZPL}, $\hbar \omega$, and the \ac{HRF} (see Table~\ref{tab:ComparisonDefectParameters}) compared to other carbon clusters~\cite{Li2023b}. Differences might originate from changes in stiffness of the environment around the clusters at the interface. However, the theoretical values consider only direct transitions of bound excitons that involve a donor state. As the experimentally observed emission peaks are symmetric, excluding significant Coulomb interaction, the \ac{DAP} transition cannot be related to a typical pair of defects, but must be related to a single complex. Due to the clear signature of the \ac{LVM}, we therefore concluded that carbon clusters probably serve as building blocks of such a defect complex that features the observed \ac{DAP} transition. Note that such a carbon cluster complex has not been investigated so far by \ac{DFT} calculations.

\begin{table}[]
\centering
\caption{\label{tab:ComparisonDefectParameters}
Comparison between defect parameters obtained by experiment and theoretically calculated values from literature~\cite{Li2023b}.}
\begin{tabular}{c|c|cccc}
Parameter & Exp. & \acs{TCAk} & \acs{TCAh} & \acs{TCIk} & \acs{TCIh} \\
\hline
\ac{ZPL} [\SI{}{\electronvolt}] & 2.53 & 2.56 & 2.67 & 2.36 & 2.49 \\
$\hbar \omega$ [\SI{}{\milli\electronvolt}] & 220 & 249 & 247 & 193 & 192 \\
$S$ & 4.83 & 2.59 & 2.35 & 5.1 & 5.5
\end{tabular}
\end{table}

\subsection{Relation Between Threshold Voltage Shift and Light Emission}
Apparently, both the blue and red components are the result of recombination, prior to which a transient shift in \ac{Vth} is created (see Fig.~\ref{fig:BandDiagram}). As shown in previous studies \cite{Feil2023,Feil2022} and also here in Section~\ref{sec:VoltageLevels}, there is a clear correlation between the transient \ac{Vth} shift and light emission for both positive and negative bias. Charging and discharging of the involved donor and acceptor states creates this correlation (see Fig.~\ref{fig:BandDiagram}).

\textit{Rising Transition:} Before a recombination event, the negative bias during the \ac{VL} phase leads to trapping of holes in the donor state, D, close to the edge of the valence band. This leads to a decreased \ac{Vth}.

\textit{Falling Transition:} Before a \ac{DAP} recombination event, the positive bias during the \ac{VH} phase leads to trapping of electrons in the acceptor state, A, close to the edge of the conduction band. This leads to an increased \ac{Vth}.

We stipulate that this is the reason for the bias level dependence, as observed in Fig.~\ref{fig:GateVoltageSweeps}, and the relation between \ac{Vth} and the light emission following equation~(\ref{eq:ICVth}). The more the donor or acceptor states get charged during the \ac{VL} and \ac{VH} phases, respectively, the more the charges undergo recombination and the higher the absolute \ac{Vth} shift prior to the switching event.

\subsection{Competing \ac{DAP} Recombination and Non-Radiative Detrapping}\label{sec:CompDAPDetrapping}

One interesting observation was that the red component increased at the rising transition with increasing rise time, and at the falling transition with increasing fall time. In charge pumping experiments, the recombination current usually decreases with increasing transition time~\cite{Groeseneken1984,Salinaro2015}, which was observed for the blue component. However, as the process of recombination depends on both donor and acceptor states, the red component can also increase with the transition time. Let us consider the falling transition as an example: During the transition, previously trapped electrons can non-radiatively detrap from the acceptor state back to the conduction band via an \ac{NMP} process (see Fig.~\ref{fig:BandDiagram}). Moreover, the holes must be trapped in the donor state close to the valence band before \ac{DAP} recombination. Depending on the time constants of the trapping and detrapping processes, which are strongly bias dependent, either an increasing or decreasing \ac{DAP} recombination with increasing transition time can be observed.

\section{Conclusion}

In our study, we conducted time-gated spectral detection of light emission through the backside of a fully-processed \ac{SiC} power \ac{MOSFET} with remaining metal drain contact during gate switching. Our findings revealed the existence of two spectral components that constituted the emission spectrum and that we referred to as the red and blue components.

Based on fitting the red component with a quantum mechanical model, the red component was identified as a \ac{DAP} recombination process involving a \ac{LVM} of \SI{220}{\milli\electronvolt}, characterized by a \ac{ZPL} and four \ac{LVM} sidebands. Such a highly energetic \ac{LVM}, well above the highest phonon modes in 4H-\ac{SiC} and \ac{SiO2}, is in general extremely rare, which allowed us to assign the red component to a carbon cluster-like defect complex at the 4H-\ac{SiC}/\ch{SiO2} interface.

The blue component could stem from two different \ac{DAP} recombination pathways. The first pathway is between the D-center and an EK$_2$ center, and the second is between nitrogen and aluminum dopants. These pathways could be identified by comparing the blue component with the emission spectrum of the body diode, which has been more thoroughly studied in literature. However, in our experiments, the blue component behaved rather like a recombination between a donor-like defect close to the valence band and channel electrons.

We were also able to link the bias level dependence of the spectral components to the transient \ac{Vth} shift and the \ac{CV} characteristic that confirms the understanding of the two components. Furthermore, we observed an independence of the emission spectra on switching frequency (\SIrange{20}{1000}{\kilo\hertz}) and duty cycle (\SIrange{5}{95}{\percent} at \SI{100}{\kilo\hertz}), revealing extremely fast trapping processes with time constants below \SI{500}{\nano\second} in strong inversion or accumulation.

In addition, we successfully traced the changes in the spectral components during a \ac{VGS} period and analyzed the effects of varying transition times. The results further confirmed the understanding of the two components and we found its linear dependence on the logarithm of the transition time, as observed for recombination current in charge pumping experiments.

Overall, our methodology of time-gated optical spectroscopy through the backside of a fully-processed \ac{SiC} \ac{MOSFET} opens new avenues for characterizing and identifying interface states and related recombination processes. This method of characterizing interface states paves the way for further research on improving the reliability and performance of these devices.

\begin{acknowledgments}
The financial support by the Austrian Federal Ministry for Digital and Economic Affairs, the National Foundation for Research, Technology and Development, and the Christian Doppler Research Association is gratefully acknowledged.
\end{acknowledgments}

\appendix

\section{Wavelength and Intensity Calibration}\label{app:Calibration}

\begin{figure}[t!]
\includegraphics[width = \columnwidth]{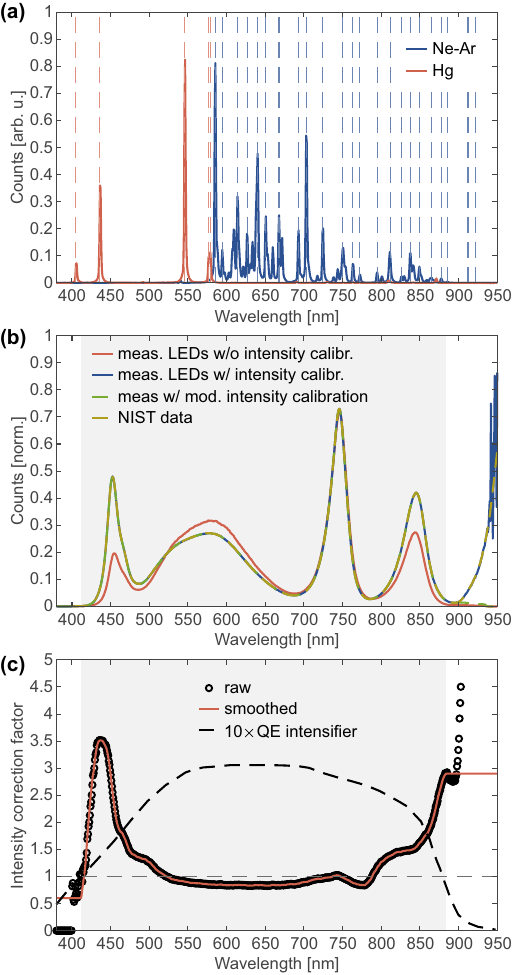}
\caption{\label{fig:Calibration} \textbf{Intensity and wavelength calibration of spectrograph and \ac{ICCD}.} \textbf{(a)}~Emission spectra of \ac{Hg} and \ac{NeAr} lamps measured with the wavelength-calibrated setup. The vertical line indicates the true position of the respective peak. \textbf{(b)}~Spectra involved in intensity calibration. \textbf{(c)}~The correction factor for intensity calibration. It is not exclusively determined by the \ac{QE} of the intensifier, but it rather reflects the efficiencies of all optical components along the light path between the entrance slit of the spectrograph and the \ac{CCD}.}
\end{figure}

To understand and model the light emission from a \ac{SiC} \ac{MOSFET}, calibration of the measurement system consisting of the spectrograph and the \ac{ICCD} is important. Also, as the investigated light emission covers the entire range of visible light, intensity and wavelength calibration are of particular importance. If either of the two is not performed, it can have strong impact on both shape and energetic position of the emission peaks. For our experiments, each spectroscopic measurement was performed with a background correction of the \ac{CCD}. Both calibrations were performed using the IntelliCal system from Teledyne Princeton Instruments.

For wavelength calibration, we used a neon-argon light source and achieved a calibration in the range of \SIrange{585}{966}{\nano\metre} with an \ac{RMS} of \SI{0.2}{\nano\metre}. We also validated the calibration using a mercury lamp with emission lines in the range of \SIrange{254}{579}{\nano\metre} (see Fig.~\ref{fig:Calibration}a) to confirm the calibration for shorter wavelengths. For intensity calibration, we used a set of temperature-compensated, \acp{LED} that provide the spectrum as shown in Fig.~\ref{fig:Calibration}b measured at the \ac{NIST}. Using IntelliCal, we accordingly performed an intensity calibration and extracted the correction factor, $CF$, that links the calibrated and non-calibrated spectra via
\begin{equation}
I_\mathrm{cal} \left( \lambda \right) = CF \left( \lambda \right) \cdot I_\mathrm{raw} \left( \lambda \right).
\end{equation}
Subsequently, we determined a wavelength range where we could particularly trust our measured spectra. This range was defined such that the measured non-calibrated intensity in each pixel, normalized to the maximum at \SI{746}{\nano\metre}, was above \SI[print-unity-mantissa=false]{1E-2}{}. The trusted wavelength range located between \SI{412}{\nano\metre} and \SI{884}{\nano\metre} is indicated by a grey background. Outside the trusted wavelength range, we set $CF$ to the last value of a smoothed (local regression with a $2^\mathrm{nd}$ degree polynomial and weighted linear least squares) within the trusted wavelength region such that $CF$ becomes constant (see Fig.~\ref{fig:Calibration}c). Note that the set of \acp{LED} provided light emission in the non-trusted region above \SI{884}{\nano\metre}. However, as the quantum efficiency of the built-in intensifier in the \ac{ICCD} strongly drops in this region, this yields a steeply increasing correction factor. The light is simply not detected. For the non-trusted region below \SI{412}{\nano\metre}, however, there is no light from the \acp{LED}, which is why the correction factor approaches zero although the quantum efficiency is still significant (confirmed by the detection of Mercury lines at \SI{405}{\nano\metre} and \SI{436}{\nano\metre}, as shown in Fig.~\ref{fig:Calibration}a). Therefore, if there was light emission below \SI{412}{\nano\metre}, we would still see the light in the raw data but not in the calibrated data. In summary, this is the reason behind defining trusted and non-trusted regions. By setting the correction factor constant in the non-trusted regions, instead of setting it down to zero, we could still observe whether we detected light in the region below \SI{412}{\nano\metre}.

\section{Open Versus Partially-Grounded Drain-Contact}

\begin{figure}[]
\includegraphics[width = \columnwidth]{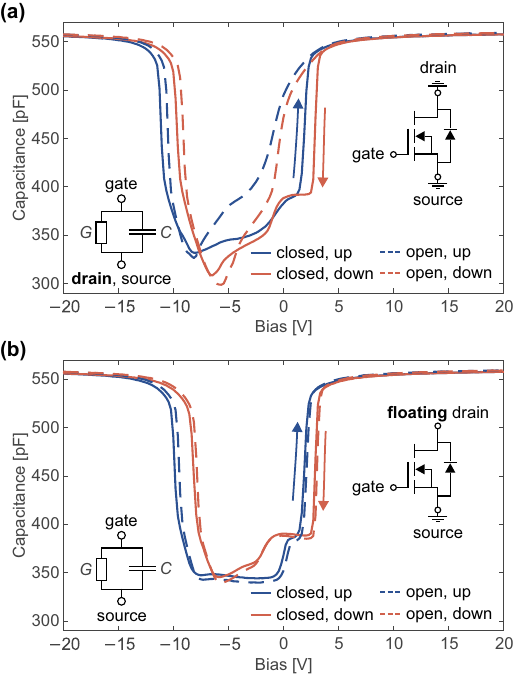}
\caption{\label{fig:CVCurves} \textbf{\Ac{CV} curves of pristine devices and devices that were opened from the backside.} \textbf{(a)}~The \ac{CV} curves of a pristine (closed) devices and an opened device with grounded drain terminals. \textbf{(b)}~The \ac{CV} curves of a pristine (closed) devices and an opened device with floating drain terminals.}
\end{figure}

There was a change in the device characteristic when the drain contact was partially removed, which is shown in Fig.~\ref{fig:CVCurves}a. However, this difference vanished almost completely when the drain terminal was kept floating, as shown in Fig.~\ref{fig:CVCurves}b. The bias dependence of the emission spectra, shown in Fig.~\ref{fig:GateVoltageSweeps}e--f, reflect this behavior.

\section{Model for the Red Component}\label{app:QMModel}
To determine the transition rate, $k_{mn}$, from the vibrational level, $m$, of electronic state $2$ to the vibrational level, $n$, of state $1$ (see Fig.~\ref{fig:ExperimentalSetup}a), Fermi's Golden rule
\begin{equation}
k_{mn} = \frac{2\pi}{\hbar} \abs{\mel{\Psi_{m}^{2}}{\hat{\mu}}{\Psi_{n}^{1}}} \delta \left( E_{m}^{2} - E_{n}^{1} - \hbar \omegaphoton \right)
\end{equation}
is used. It is a consequence of time-dependent perturbation theory. Hereby, $\hat{\mu}$ is the perturbation operator, which in the case of photon emission is the dipole operator.

Using the Born-Oppenheimer approximation~\cite{Schwartz1973}, the total wave function can be expressed as a product of the electronic wave function $\psi$, the rotational wave function $\Phi$, and the vibrational wave function $\phi$
\begin{equation}
\Psi_{n}^{i} = \underbrace{\psi_{~}^{i}}_{\text{electronic}} \cdot \underbrace{\Phi^{i} \phi_{n}^{i}}_{\text{nuclear}}.
\end{equation}
The dipole operator is assumed to be independent of the nuclear coordinates, which allows to consider only the vibrational component (Franck-Condon factor). Additionally, a factor of $\omegaphoton^3$ enters the equation~\cite{Razinkovas2021}.

\begin{equation}
k_{mn} \propto \omegaphoton^3 \cdot \abs{\bra{\phi_{m}^{2}}\ket{\phi_{n}^{1}}}^2 \delta \left( E_{m}^{2} - E_{n}^{1} - \hbar \omegaphoton \right)
\label{eq:TransitionRateSinglePeak}
\end{equation}

For each of the two states, we approximated the Born-Oppenheimer surfaces by harmonic potentials in one dimension, leading to the well-known Schrödinger equation
\begin{equation}
-\frac{\hbar^2}{2m_i} \frac{\mathrm{d}^2}{\mathrm{d}x^2} \phi_{n}^{i} \left( x \right) + \frac{1}{2} m_i \omega_i^2 x^2 \phi_{n}^{i} \left( x \right) = E_{n}^{i} \phi_{n}^{i} \left( x \right)
\end{equation}
that has to be solved for each of the two states. Depending on the considered energy range, a description solely based on a harmonic potential may not always be sufficient, because the underlying Born-Oppenheimer potential could be anharmonic~\cite{Yan2012}, which leads to a gradually decreasing spacing in the vibrational energy levels with increasing energy. This could be modelled by replacing the harmonic potential by the Morse potential~\cite{McCluskey2000,Morse1929}. However, as we observed equally spaced emission peaks within the considered spectral range, the harmonic potential was fully sufficient. With the coordinate transformation
\begin{equation}\label{eq:CoordinateTrans}
x \rightarrow q = \sqrt{\frac{m_i \omega_i}{\hbar}} x
\end{equation}
the Schrödinger equation simplifies to
\begin{equation}
-\frac{1}{2} \hbar \omega_i \frac{\mathrm{d}^2}{\mathrm{d}q^2} \phi_{n}^{i} \left( q \right) + \frac{1}{2} \hbar \omega_i q^2 \phi_{n}^{i} \left( x \right) = E_{n}^{i} \phi_{n}^{i} \left( q \right),
\end{equation}
whereby the potential translates to
\begin{equation}
V \left( q \right) = \frac{1}{2} \hbar \omega_i q^2.
\end{equation}
The solution of the Schrödinger equation of the harmonic oscillator can be found in any textbook on basic quantum mechanics. Hereby, the vibrational eigenstates $\phi_{n}^{i}$ of electronic state $i \in \{1, 2\}$, numbered by index $n \in \{0, 1, 2, 3, ...\}$, are given by
\begin{equation}
\phi_{n}^{i} \left( q \right) = \frac{1}{\left( 2^n n! \right)} \left( \frac{1}{\pi} \right)^\frac{1}{4} e^{-\frac{q^2}{2}} H_n \left( q \right)
\end{equation}
with $H_n$ being the $n^\text{th}$ Hermite polynomial. Furthermore, the energy eigenvalues $E_{n}^{i}$ are given by
\begin{equation}
E_{n}^{i} = \hbar \omega_i \left( \frac{1}{2} + n \right).
\end{equation}
Consequently, $\hbar \omega$ is the distance between the energy levels and scales with the curvature of the harmonic potential. The potentials of states $1$ and $2$ are separated by $\Delta q$ and $\dETwoOne$.

In the most general case, the pure underlying lifetime-limited lineshape is a Laurentzian following
\begin{equation}
f \left( \Ephoton, E_{mn}, \dELor \right) = \frac{\dELor/\left(2 \pi \right)}{\left( \Ephoton - E_{mn} \right)^2 + \left( \dELor/2 \right)^2}.
\end{equation}
Broadening of this lineshape was introduced by the assumption of a Gaussian distribution in $\omega_1$, $\omega_2$ and $\dETwoOne$. As convolution operations follow commutativity and associativity and a convolution of two Gaussians yields another Gaussian with a modified width, we can compute the final lineshape as a convolution of the Laurentzian and a Gaussian.
\begin{align}
&L \left( \Ephoton, E_{mn}, \sigma_{mn}, \dELor \right) = \\ &\int_{-\infty}^{\infty} G \left( E', \sigma_{mn} \right) f \left( \Ephoton - E', E_{mn}, \dELor \right) \mathrm{d}E^\prime
\end{align}
Hereby, the width of the Gaussian is given by
\begin{equation}
\sigma_{mn} = \sqrt{\sigma_{1}\left( n \right)^2 + \sigma_{2}\left( m \right)^2 + \sigma_{\Delta E}^2}
\end{equation}
with standard deviations
\begin{equation}
\sigma_{i} \left( n \right) = \sigma_i \cdot \left( \frac{1}{2}+n \right)
\end{equation}
that depend on the vibrational quantum number because
\begin{align}
\Ephoton &= E_m^{2} - E_n^{1} + \dETwoOne \\
&= \hbar \omega_2 \left( \frac{1}{2} + m \right) - \hbar \omega_1 \left( \frac{1}{2} + n \right) + \dETwoOne.
\end{align}
The lineshape function, $L$, finally replaced the Dirac delta distribution in equation~(\ref{eq:TransitionRateSinglePeak}) and all considered transitions were summed up, which resulted in the emission spectrum presented in equation~(\ref{eq:ModelEmissionSpectrum}).

Finally, the \ac{HRF} $S$ is defined as the number of vibrational quanta that are involved in the radiative transition. Using the coordinate $q$ as defined in equation~(\ref{eq:CoordinateTrans}), it is therefore given by
\begin{equation}
S = \frac{1}{2} \left( \Delta q \right)^2.
\end{equation}

\end{document}